\documentclass{emulateapj}

\begin{document}

\title{Far Infrared Imaging of NGC~55}

\author{C. W. Engelbracht\altaffilmark{1}, K. D. Gordon\altaffilmark{1}, G. J.
Bendo\altaffilmark{1}, P. G.  P\'{e}rez-Gonz\'{a}lez\altaffilmark{1}, K. A.
Misselt\altaffilmark{1}, G. H.  Rieke\altaffilmark{1}, E. T.
Young\altaffilmark{1}, D. C.  Hines\altaffilmark{1,3}, D. M.
Kelly\altaffilmark{1}, J. A.  Stansberry\altaffilmark{1}, C.
Papovich\altaffilmark{1}, J. E.  Morrison\altaffilmark{1}, E.
Egami\altaffilmark{1}, K. Y. L.  Su\altaffilmark{1}, J.
Muzerolle\altaffilmark{1}, H. Dole\altaffilmark{1}, A.
Alonso-Herrero\altaffilmark{1}, J.  L. Hinz\altaffilmark{1}, P. S.
Smith\altaffilmark{1}, W. B. Latter\altaffilmark{2}, A.
Noriega-Crespo\altaffilmark{2}, D. L.  Padgett\altaffilmark{2}, J.
Rho\altaffilmark{2}, D. T. Frayer\altaffilmark{2}, and S.
Wachter\altaffilmark{2}}

\altaffiltext{1}{Steward Observatory, University of Arizona, Tucson, AZ 85721}
\altaffiltext{2}{Spitzer Science Center, California Institute of Technology,
Pasadena, CA 91125}
\altaffiltext{3}{Space Science Institute, 4750 Walnut Street, Suite 205,
Boulder, CO 80301}

\slugcomment{To appear in ApJS Spitzer Special Issue}

\begin{abstract}

We present images of the galaxy NGC~55 at 24, 70, and 160~\micron\ obtained
with the Multiband Imaging Photometer for Spitzer (MIPS) instrument aboard the
{\it Spitzer Space Telescope}.  The new images display the far infrared
emission in unprecedented detail and demonstrate that the infrared morphology
differs dramatically from that at shorter wavelengths.  The most luminous
emission region in the galaxy is marginally resolved at 24\micron\ and has a
projected separation of nearly 520~pc from the peak emission in the optical
and near infrared.  This region is responsible for $\sim9$\% of the total
emission at 24~\micron\ and is likely a young star formation region.  We show
that this and other compact sources account for more than $1/3$ of the total
24~\micron\ emission.  We compute a total infrared luminosity for NGC~55 of
$1.2\times10^9~L_\odot$.  The star formation rate implied by our measurements
is $0.22~M_\odot~yr^{-1}$.  We demonstrate that the cold dust is more extended
than the warm dust in NGC~55---the minor-axis scale heights are 0.32, 0.43,
and 0.49~kpc at 24, 70 and 160~\micron, respectively.  The dust temperature
map shows a range of temperatures that are well-correlated with the
24~\micron\ surface brightness, from 20~K in low-surface-brightness regions to
26~K in high-surface-brightness regions.

\end{abstract}


\section{Introduction}

NGC~55 is a dwarf ($M_B = -18.6$) member of the Sculptor group, a nearby group
dominated by spiral galaxies such as NGC~300 and the famous starburst NGC~253.
At a distance of 1.78~Mpc \citep{kar03}, the high inclination of NGC~55
\citep[81\arcdeg;][]{kis88} makes it the nearest bright ($m_B < 9$), edge-on
galaxy.  The galaxy is an analogue of the Large Magellanic Cloud, being of
similar luminosity, morphological type (SB(s)m), and metallicity
\citep[$12 + {\rm log} (O/H) = 8.33$;][]{sch93}.

The energetic star formation, proximity, and high surface brightness of NGC~55
have made it a popular target for ISM studies, both of the neutral
\citep[e.g.,][]{puc91} and ionized component \citep[e.g.,][]{hoo96,fer96}.
The galaxy is known to host extraordinary \ion{H}{2} regions both in
\citep{tul04} and out of the disk \citep{tul03}.

NGC 55 was observed as part of the Multiband Imaging Photometer for Spitzer
(MIPS) instrument commissioning activities \citep{rie04}.  NGC~55 has a large
($\sim$30\arcmin) but well-defined angular extent and so was chosen to test
the imaging capabilities of the instrument on extended sources with both high
and low surface brightness regions.  These commissioning observations proved
to be of sufficient quality to support scientific analysis.  In this paper, we
measure global properties of the galaxy (temperature, luminosity, scale height
as a function of wavelength, and star formation rate) and exploit the
sensitivity and resolution of MIPS to determine the distribution of
temperatures and luminosities within the galaxy and to address the nature of
the nuclear source.  We refer to other papers in this volume for topics best
studied in more face-on galaxies:  1.) Star formation rate indicators in M~81
\citep{gor04b}; 2.) Location of stochastically heated PAH dust grains in
NGC~300 \citep{hel04}; 3.) Heating of cold dust in M~33 \citep{hin04}.

\section{Observations and Data Reduction}
\label{sec:data}

Images of NGC~55 at 24, 70, and 160~\micron\ were obtained with MIPS on 2003
November 30 (PROGID = 718, AORKEY = 8085504, 8100352).  3232 individual images
at each of the three MIPS bands were taken, with an exposure time of
approximately 4 seconds per image.  The depth of the map at each point was
240, 120, and 24 seconds at 24, 70, and 160~\micron, respectively.  The
resulting mosaics cover an area of approximately $30\arcmin \times 60\arcmin$
at each wavelength, with a region in common to all three wavelengths of
$30\arcmin \times 30\arcmin$.

\begin{figure}
\figurenum{1}
\epsscale{1.2}
\plotone{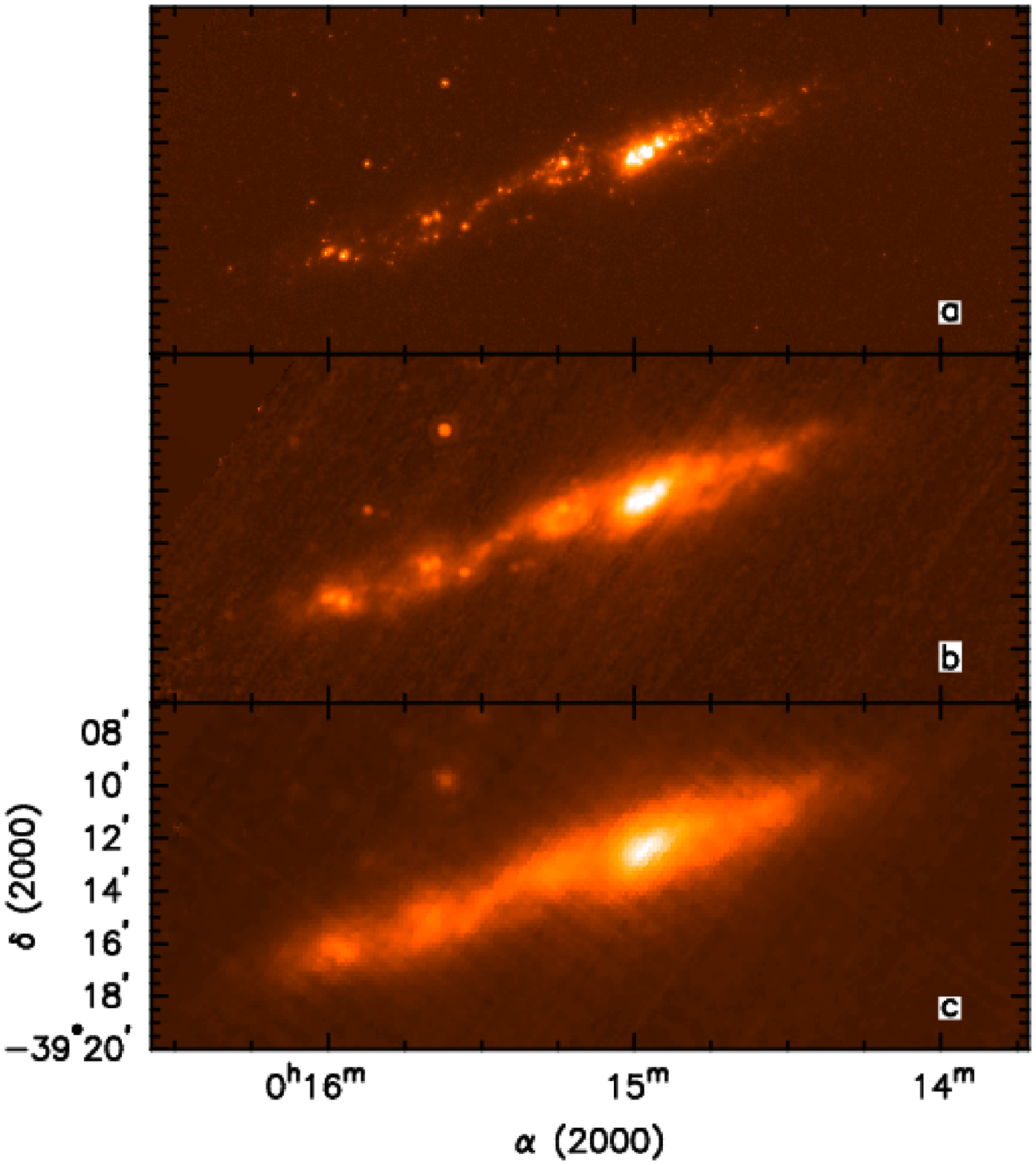}
\caption{Images of NGC~55 at 24~\micron\ (a), 70~\micron\ (b), and 160~\micron\
(c), presented in an asinh stretch to compress the dynamic range.}
\label{fig:images}
\end{figure}

The MIPS images were reduced using the MIPS Instrument Team Data Analysis Tool
which is the test bed for data reduction algorithms for MIPS \citep{gor04a}.
The data were reduced as described in that paper, with some minor exceptions:
a customized flat field was required at 24~\micron, and a time-dependent
illumination correction and calibration lamp latent image removal were
required at 70~\micron.  The uncertainties on the calibration are estimated at
10\%, 20\%, and 20\% for 24, 70, and 160~\micron, respectively \citep[cf.
][]{rie04}.  These uncertainties are dominated by the uncertainty in the
absolute calibration---uncertainties in the background level for this target
are around 5\% and so have not been included in subsequent analysis.  The
final MIPS mosaics are displayed in Fig.~\ref{fig:images}.

The beam sizes in the three MIPS bands are limited by diffraction and so vary
with wavelength:  The FWHM of the point-spread-function (PSF) is 6, 18, and
40\arcsec\ at 24, 70, and 160~\micron, respectively.  Where images have been
compared directly at different wavelengths, they have been convolved to
matching resolutions (the 160\micron\ resolution, unless otherwise noted)
using kernels derived from PSF models generated using the TinyTim software
\citep{kri02}.

\section{Results}

\subsection{Morphology}
\label{sec:morphology}

\begin{figure}
\figurenum{2}
\epsscale{1.2}
\plotone{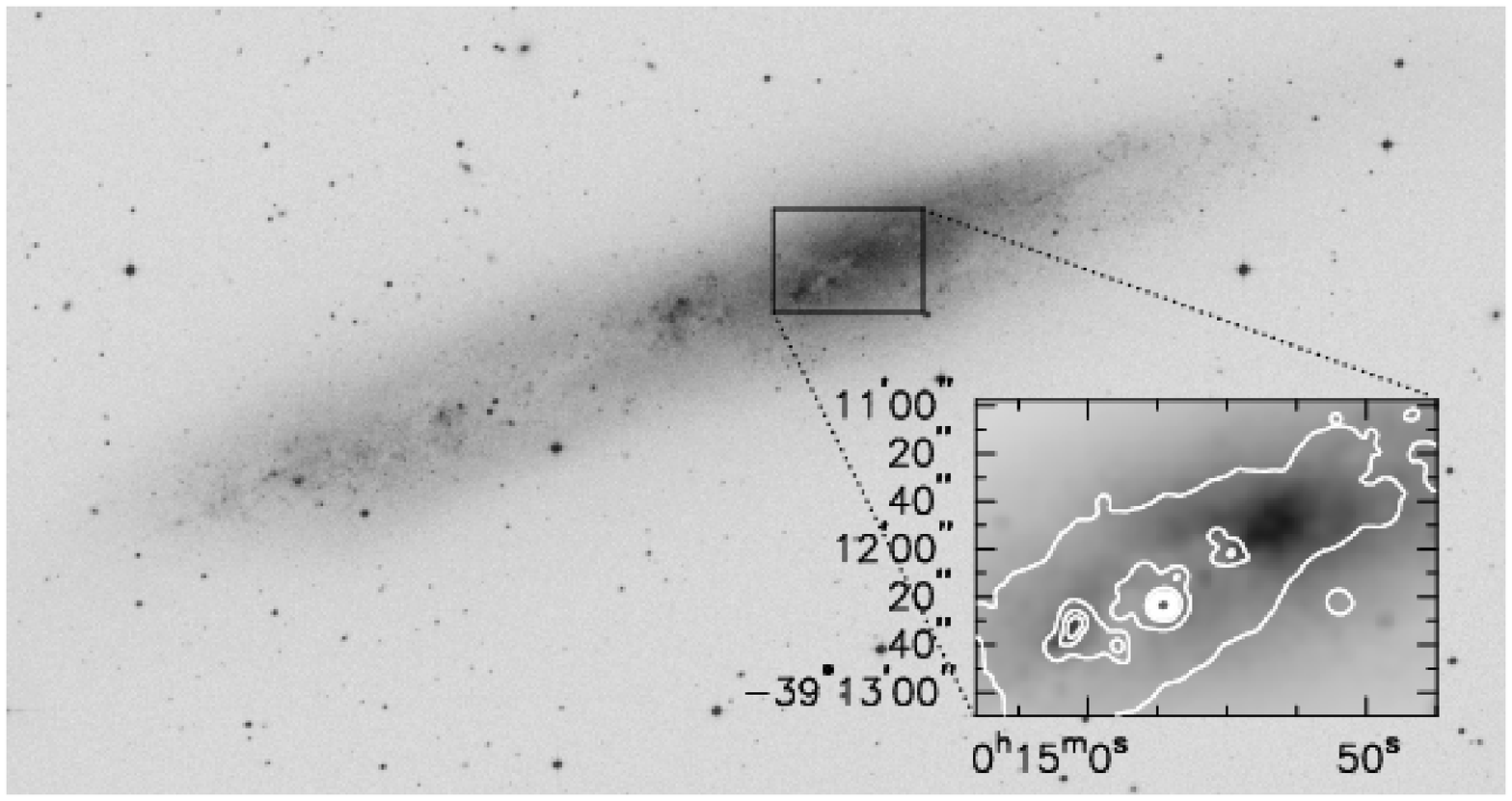}
\caption{Digitized Sky Survey image of NGC~55.  The inset shows the I band
image (obtained from the NASA/IPAC Extragalactic Database and attributed there
to B. Madore) in a linear stretch.  Plotted over the I band image are
24~\micron\ contours in evenly-spaced intervals from 50 to
4000~$\mu$Jy~arcsec.$^{-2}$.  The I band image has been convolved to match the
24\micron\ resolution.}
\label{fig:optical}
\end{figure}

The morphology of the galaxy in the MIPS images is broadly similar, being
dominated by a bright central region and asymmetric emission from the disk on
either side.  The extent of the galaxy (both in the optical and far infrared
(FIR)) is $\sim9$\arcmin\ to the NW of the nuclear region and $\sim15$\arcmin\
to the SE.  The central region and the region to the SE are separated by a
low-surface-brightness gap that crosses the galaxy at
$\alpha(2000)\sim00^h15^m08^s$.  The 24~\micron\ emission is dominated by
discrete regions, in contrast to the smoother structure seen at 70 and
160~\micron.  These discrete emission regions also dominate the nuclear region
(which we have assumed lies at the 24\micron\ peak), as demonstrated by
Figure~\ref{fig:optical}.

\subsection{Discrete Sources}
\label{sec:discrete}

\begin{figure}
\figurenum{3}
\epsscale{1.2}
\plotone{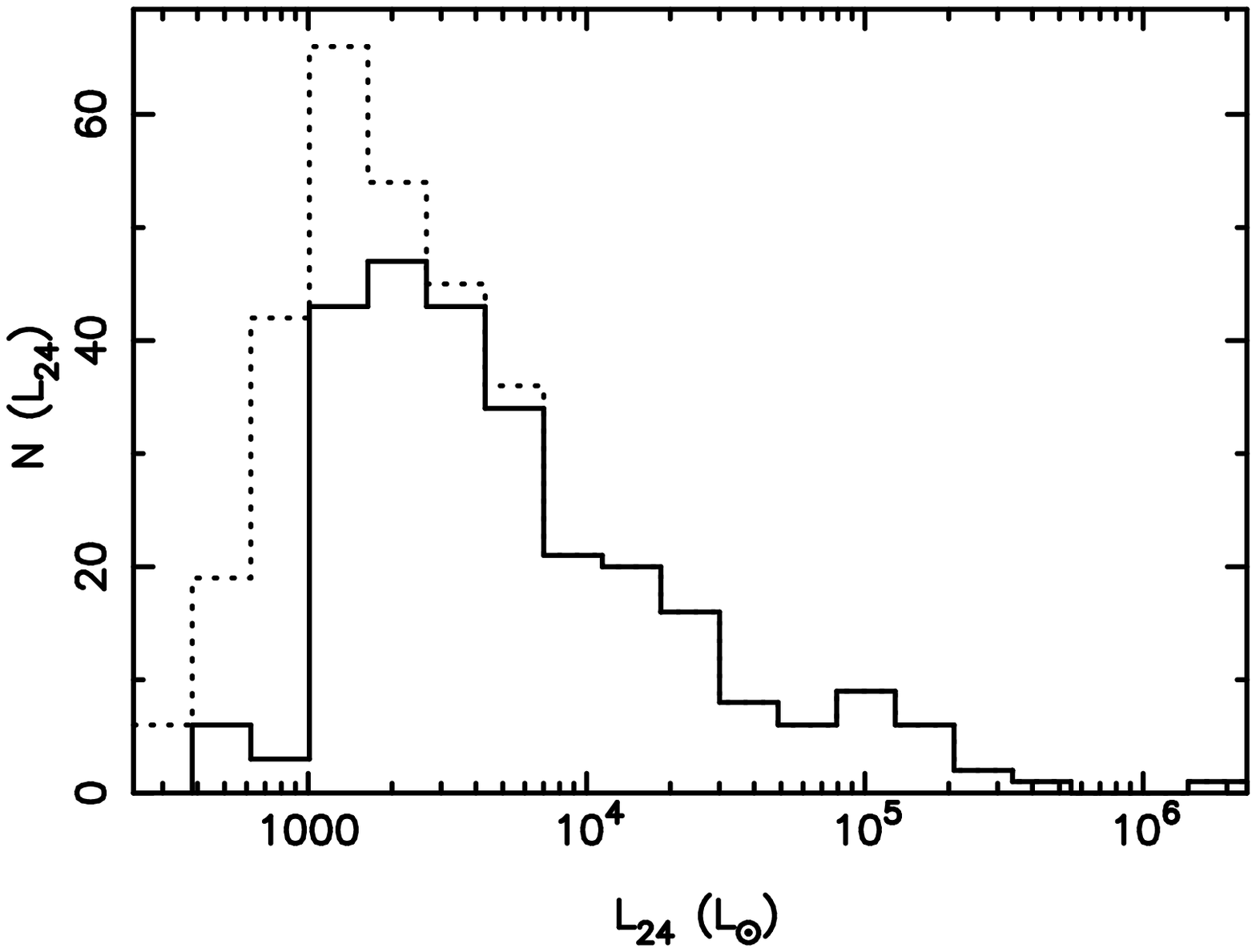}
\caption{Luminosity distribution of discrete 24~\micron\ emission regions.
The dotted line represents the raw histogram while the solid line represents
the luminosity function after correction for background sources.}
\label{fig:lum24}
\end{figure}

We performed photometry on the 24~\micron\ image using the DAOPhot package in
IRAF\footnote{IRAF is distributed by the National Optical Astronomy
Observatories, which are operated by the Association of Universities for
Research in Astronomy, Inc., under cooperative agreement with the National
Science Foundation.}.  The luminosities of the extracted point sources are
plotted in Figure~\ref{fig:lum24}, where we simply multiplied the measured
flux density by the 24~\micron\ bandpass to determine the in-band flux and
thence the luminosity.  We have made a correction for the background point
sources in the NGC~55 image by subtracting the luminosity function measured in
a background region of identical size, well away from the galaxy.  The
integrated flux of the extracted point sources is more than $1/3$ of the total
24~\micron\ emission from this galaxy.

The brightest point source is located near the center of NGC~55.  As
demonstrated by Figure~\ref{fig:optical}, this 24~\micron\ peak is nearly an
arcminute (a projected separation of 520~pc) away from the peak I-band
emission.  This source appears to be hidden behind a strong dust lane apparent
in the I-band image and is responsible for $\sim9$\% of the total emission at
24~\micron.  This source is marginally resolved at 24~\micron---the measured
FWHM of the source is $\sim$5\farcs3.

\subsection{Major and Minor Axis Profiles}

\begin{figure}
\figurenum{4}
\epsscale{1.2}
\plotone{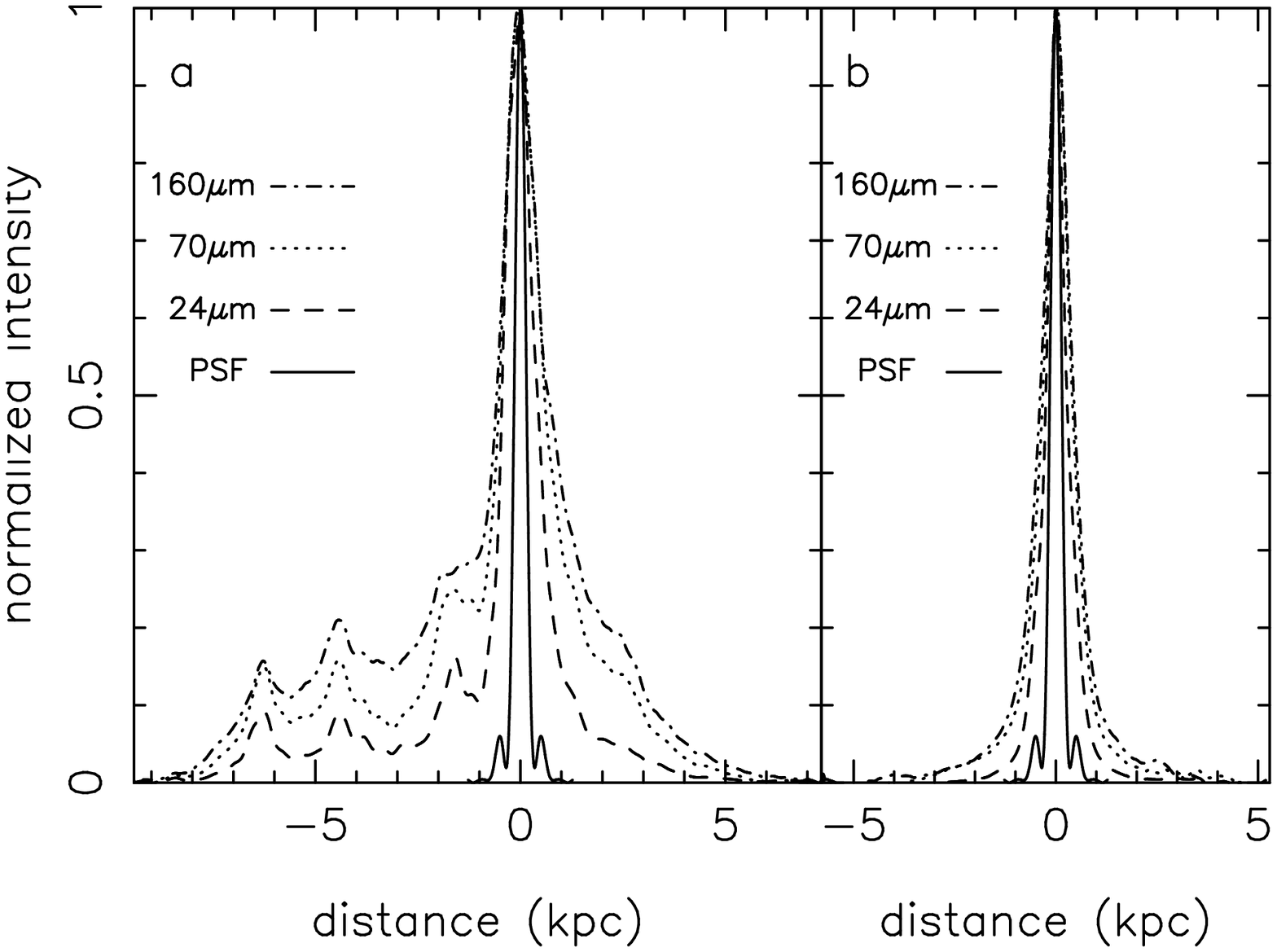}
\caption{Surface brightness profiles along the major (a) and minor (b) axes of
NGC~55, plotted on a linear scale.  The major-axis profile runs E to W along
the disk while the minor-axis profile runs S to N across the disk.}
\label{fig:profiles}
\end{figure}

The edge-on aspect of the galaxy provides a unique opportunity to measure
infrared scale heights.  We generated profiles along the major and minor axes
of NGC~55 by averaging the images (convolved to the 160\micron\ resolution)
parallel and perpendicular to the disk of the galaxy.  These profiles
are presented in Figure~\ref{fig:profiles}, in which the PSF measured on an
unresolved galaxy in the field is plotted for reference.  The profiles have
all been normalized to the same peak value.

The extent of the observed emission is correlated with wavelength:  the
emission is more extended at longer wavelengths, along both the major and
minor axes.  The major-axis profile is highly asymmetric, with an off-center
nuclear region and discrete emission regions evident in the eastern portion of
the disk.  The minor-axis profile is more symmetric, and we fit it with a
function proportional to $exp (-x/h)$, where x is the distance from the
emission peak and h is the scale height.  The scale height is 0.32, 0.43, and
0.49~kpc at 24, 70 and 160~\micron.  These values are higher than the value of
$0.10-0.21$~kpc determined by \citet{kis88} for the stellar component of
NGC~55 but are comparable to the $0.37-0.45$~kpc scale height for the ionized
gas determined by \citet{mil03}.

\subsection{Spectral Energy Distribution}
\label{sec:sed}

\begin{figure}
\figurenum{5}
\epsscale{1.2}
\plotone{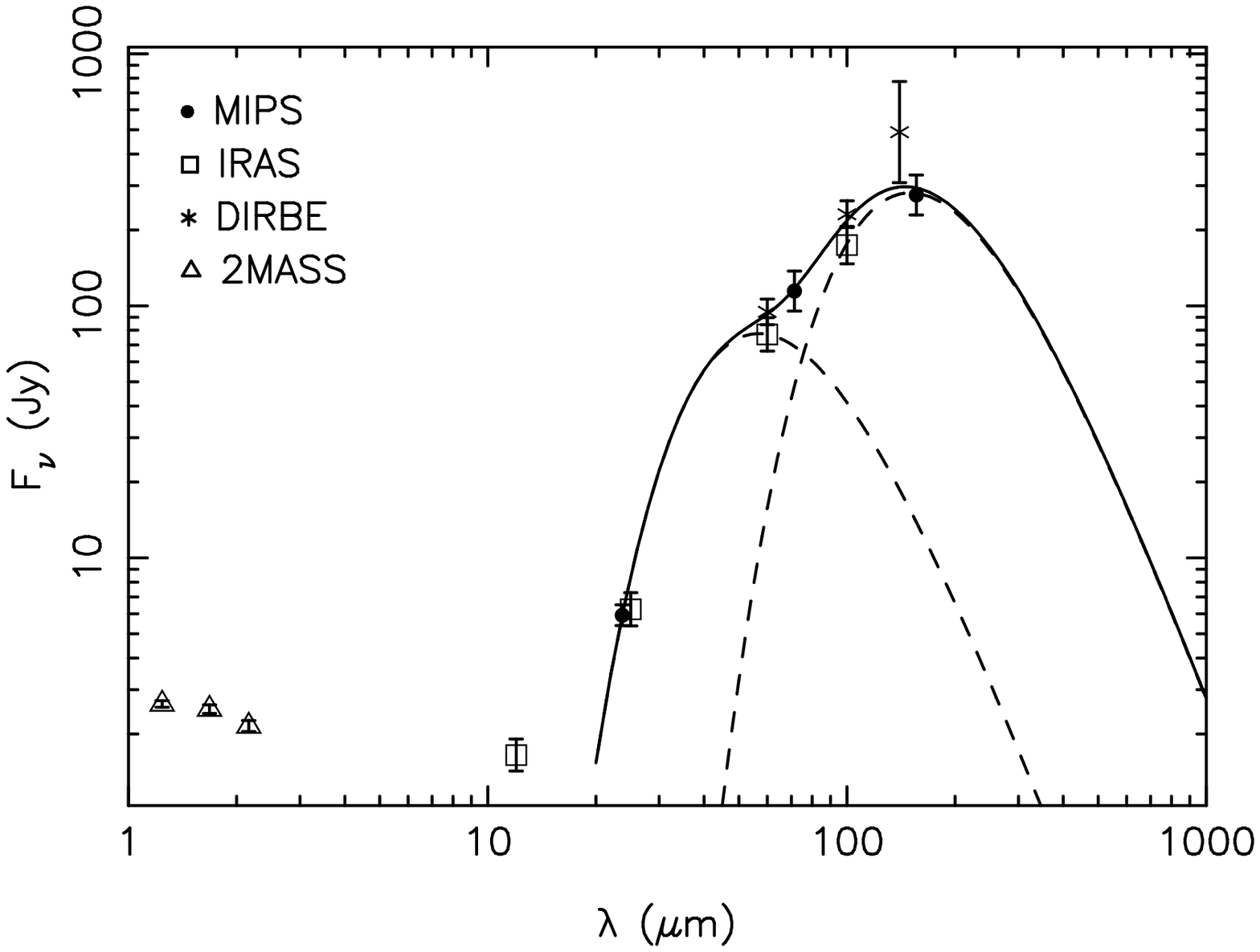}
\caption{Infrared SED of NGC~55.  The dashed lines are modified blackbody
curves (emissivity $\sim\lambda^{-2}$) at 51~K and 19~K, while the solid line
is the sum of the two dashed curves.}
\label{fig:sed}
\end{figure}

We computed the FIR SED of NGC~55 using MIPS data, combined with measurements
by the Infrared Astronomical Satellite (IRAS), the Diffuse Infrared Background
Experiment (DIRBE), and the Two Micron All Sky Survey (2MASS).  The total
emission in the MIPS images was measured using IRAF's ``polyphot'' task, while
the IRAS measurements were taken from \citet{ric88}.  The DIRBE measurements
were computed using the DIRBE Point Source Photometry Research
Tool\footnote{\url{http://lambda.gsfc.nasa.gov/product/cobe/browser.cfm}}, and
the 2MASS measurements were taken from \citet{jar03}.  We fit the fluxes with
two modified blackbody functions to characterize roughly the dust temperature
and found a good fit with two components at 51~K and 19~K, modified by a
$\lambda^{-2}$ emissivity \citep[adopted as an appropriate factor beyond
70~\micron; cf.][]{dra03}.  The data and the fit are plotted in
Figure~\ref{fig:sed}.

The total infrared (TIR) luminosity was estimated from both the MIPS and IRAS
fluxes using equations 4 and 5 of \citet{dal02}, respectively:  both equations
indicate the luminosity of NGC~55 is $1.14\times10^9~L_\odot$.  Additionally,
the bolometric IR luminosity was computed from the data directly, integrating
the two-temperature blackbody fit discussed above (plus a linear fit to the
2MASS and 12~\micron\ data where the blackbody falls short of the observed
emission) from 3-1100~\micron\ to compute a luminosity of
$1.29\times10^9~L_\odot$.  We have averaged these three estimates to compute
$L_{TIR}=1.2\times10^9~L_\odot$.  The star formation rate implied by this
luminosity is $0.22~M_\odot~yr^{-1}$, using equation~4 of \citet{ken98}.  This
rate is roughly 40\% higher than the value derived from H$\alpha$ measurements
\citep{fer96}.

We compute a total infrared luminosity of $1.2\times10^8~L_\odot$ for the
brightest compact source discussed in \S~\ref{sec:discrete} using the formula
discussed above after making an aperture correction of 2.5 (computed using the
PSF model discussed in \S~\ref{sec:data}).  This value depends on a large
aperture correction, so we have compared it to the value obtained by assuming
that the SED of the integrated galaxy is similar to that of the compact source
(measured at the native 24~\micron\ resolution) and that we can therefore
scale the 24~\micron\ emission to a TIR luminosity.  Since this source
contributes $\sim9$\% of the flux at 24~\micron, the integrated luminosity
implied by taking this same fraction of $L_{TIR}$ is $1.1\times10^8~L_\odot$,
similar to the value derived above.  This large luminosity is similar in scale
to the nuclear star formation regions in M~33 and the Galactic Center
discussed by \citet{gor99}, and we suggest that the nuclear source in NGC~55
is similarly powered by star formation.  The high infrared and H$\alpha$
luminosity ($9\times10^{39}$~erg~s$^{-1}$ \citep{fer96}) of the central
source can be plausibly attributed to star formation if we are observing a
very young burst:  the output is similar to a Starburst99 \citep{lei99} model
(${\rm M} = 10^5~{\rm M}_\odot, \alpha = 2.35, Z = 0.004$) at an age of 2~Myr.
The star formation scenario for the central source is also consistent with
mid-infrared spectroscopic \citep{tho00}, radio \citep{hum86}, and x-ray
\citep{ran03} observations of this galaxy.

\subsection{Dust Content}
\label{sec:dust}

\begin{figure}
\figurenum{6}
\epsscale{1.2}
\plotone{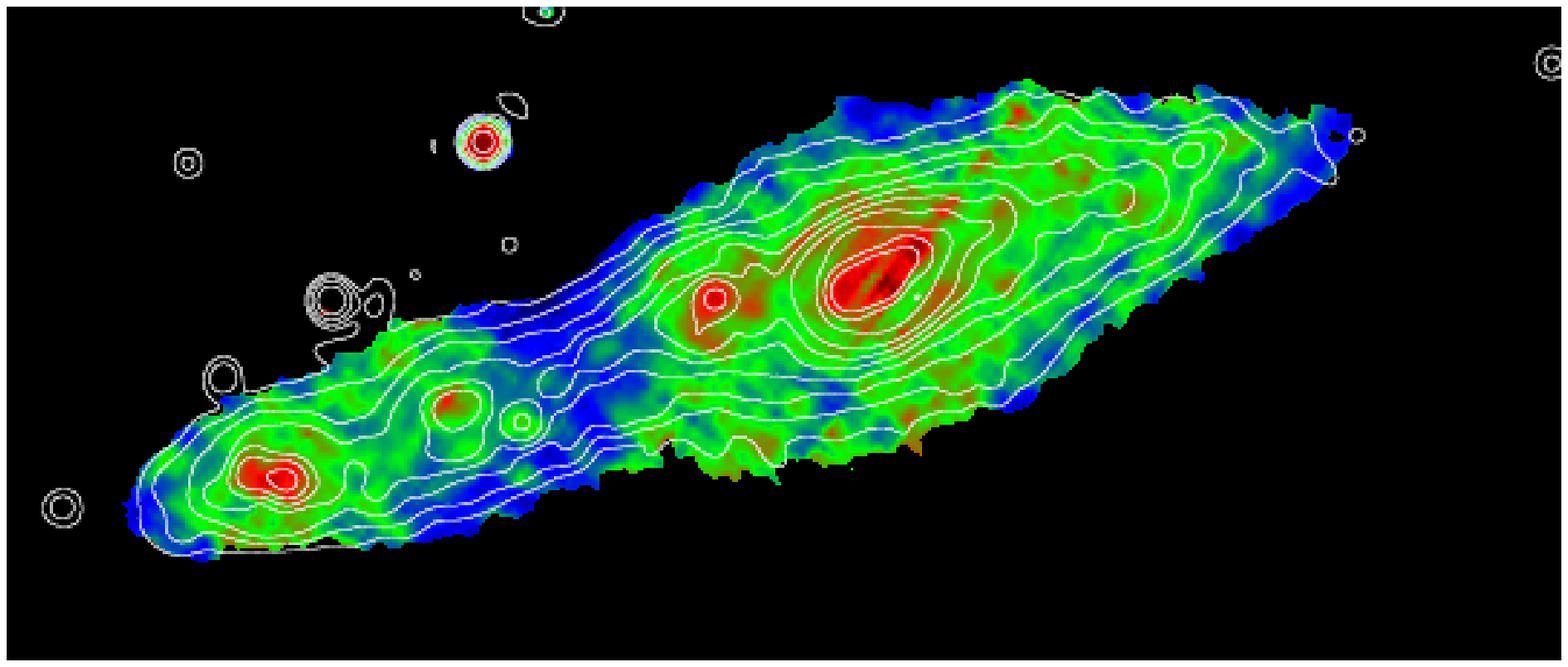}
\caption{Temperature map of NGC~55.  The color stretch ranges from 19~K (blue)
to 27~K (red), and the image has been clipped to exclude regions of low
signal-to-noise ratio in the 70 and 160~\micron\ bands.  The contours show
the 24~\micron\ emission, convolved to the 160~\micron\ resolution as
described in the text, in 10 logarithmically-spaced levels from 1.5 to 200
$\mu$Jy~arcsec.$^{-2}$.}
\label{fig:temperature}
\end{figure}

We have used the 70 and 160~\micron\ images to characterize the dust
temperature in NGC~55.  We fit a blackbody modified by a $\lambda^{-2}$
emissivity function to each pixel and assigned a temperature to that pixel
based on the temperature of the best fit.  In reality, multiple dust
temperature components are expected to contribute to the far-infrared emission
\citep{dal02}, but this approach allows us to characterize variations across
the disk of the galaxy.  The result is shown in Figure~\ref{fig:temperature}.
The mean temperature of this map (22.6~K) is similar to the value of 22~K
computed by \citet{ben03} from the central area covered by the single ISO 60
and 100~\micron\ pointing.

The temperatures range from 20 to 26~K and correlate well with the FIR surface
brightness, with the exception of a few streaks in the temperature map which
correspond to artifacts in the 70 and 160~\micron\ images.  The highest dust
temperatures in Figure~\ref{fig:temperature} correspond to peaks in the
24~\micron\ emission and are likely \ion{H}{2} regions \citep[cf.
][]{hin04,gor04b} while the lowest temperatures correspond to diffuse regions
with little star formation.

\section{Conclusion}
\label{sec:conclusion}

We have presented observations of the galaxy NGC~55 at 24, 70, and
160~\micron\ made with the MIPS instrument aboard the {\it Spitzer Space
Telescope.} The data provide images at unprecedented sensitivity and
resolution at these wavelengths.  The data demonstrate that the brightest
region at FIR wavelengths is faint at optical wavelengths and has a projected
separation from the optical peak of nearly 520~pc.  The FIR peak is barely
resolved in the 6\arcsec\ 24~\micron\ beam but still accounts for $\sim9$\% of
the total flux at that wavelength.  Comparison of the infrared luminosity of
this source and H$\alpha$ measurements from the literature to a starburst
model suggest that the current episode of star formation could be as young as
2~Myr.

The TIR luminosity of the galaxy measured in several ways is
$1.2\times10^9~L_\odot$, implying a star formation rate of
$0.22~M_\odot~yr^{-1}$.  10\% of the infrared luminosity is due to the compact
nuclear source, which is of comparable luminosity to the bright star formation
regions at the center of M~33 and the Milky Way.  The integrated flux of this
source and the other compact sources in the galaxy accounts for more than
$1/3$ of the 24~\micron\ emission from NGC~55.  A dust temperature map shows
that the temperature ranges from 20~K in the low-surface-brightness region SE
of the nucleus to 26~K in the nucleus and other star-formation regions in the
disk.

\acknowledgments

The authors wish to thank M. Blaylock and J. Cadien for their expert
assistance with processing the MIPS data presented here.

This work is based in part on observations made with the Spitzer
Space Telescope, which is operated by the Jet Propulsion Laboratory,
California Institute of Technology under NASA contract 1407. Support for
this work was provided by NASA through Contract Number 960785 issued by
JPL/Caltech.

This research has made use of the the NASA/IPAC Extragalactic Database (NED)
and the NASA/IPAC Infrared Science Archive, which are operated by the Jet
Propulsion Laboratory, California Institute of Technology, under contract with
the National Aeronautics and Space Administration.

This research used the DIRBE Point Source Photometry Research Tool, a service
provided by the Legacy Archive for Microwave Background Data at NASA's Goddard
Space Flight Center.

\clearpage

\end{document}